\documentclass[preprint,showpacs,amsmath,amssymb,nofootinbib]{revtex4}

\usepackage{dcolumn}
\usepackage{bm}

\usepackage{graphicx}
\usepackage{epstopdf}
\usepackage{pslatex}

\newcommand\be{\begin{equation}}
\newcommand\ee{\end{equation}}
\newcommand\ba{\begin{eqnarray}}
\newcommand\ea{\end{eqnarray}}
\newcommand\eq{\begin{equation}}           
\newcommand\en{\end{equation}}

\def\tkd{$T_{kd}$ }
\def\Sun{\odot}

\input epsf
\usepackage{amssymb}
\include{epsf,epsfig}

\usepackage{subfigure}

\begin{document}
\title{
The effect of quark interactions on dark matter kinetic decoupling and the mass of the smallest dark halos}
\author{Paolo Gondolo$^{1,2}$, Junji Hisano$^3$ and Kenji Kadota$^3$\\
 {\em \small
 $^1$ Department of Physics and Astronomy, University of Utah, Salt Lake City, UT 84112-0830, USA}\\
 {\em \small
 $^2$ Department of Physics and Astronomy, Seoul National University, Seoul, Korea 151-747}\\
{\em \small 
$^3$ Department of Physics, Nagoya University, Nagoya 464-8602, Japan} 
}
\begin{abstract}
The kinetic decoupling of dark matter (DM) from the primordial plasma sets the size of the first and smallest dark matter halos.
Studies of the DM kinetic decoupling have hitherto mostly neglected interactions between the DM and the quarks in the  plasma. Here we illustrate their importance using two frameworks: a version of the Minimal Supersymmetric Standard Model (MSSM) and an effective field theory with effective DM-quark interaction operators. We connect particle physics and astrophysics obtaining  bounds on the smallest dark matter halo size from collider data and from direct dark matter search experiments. In the MSSM framework,  adding DM-quark interactions to DM-lepton interactions more than doubles the smallest dark matter halo mass in a wide range of the supersymmetric parameter space.
\end{abstract}
\maketitle   
\setcounter{footnote}{0} 
\setcounter{page}{1}
\setcounter{section}{0} \setcounter{subsection}{0}
\setcounter{subsubsection}{0}

\section{Introduction}
The nature of dark matter is still an open question, despite the growing evidence in support of its existence. Weakly interacting massive particles (WIMPs) are among the favorite candidates for dark matter. One of the unique features of WIMPs is that due to their weak interaction and heavy mass, 
they can lose thermal contact with the heat bath at a relatively early stage in the history of the Universe. 
The thermal, or as often called kinetic, decoupling of dark matter from the relativistic plasma sets the scale of the smallest dark matter halos, which are the first to form. Kinetic decoupling could provide a powerful cosmological probe on the properties of dark matter, in a way analogous to baryon decoupling, which has been unveiling the nature of our Universe through the baryon acoustic oscillations and the cosmic microwave background. Kinetic decoupling has duly received a fair amount of attention, being a compelling interdisciplinary avenue to explore from both particle physics and astrophysics perspectives \cite{wider,chen,wimpy,hofm,kam2,brifirst,bert,kasahara,bri5,torsten3,stefano,bi,matias,boe,bere}. 


Kinetic decoupling is a distinct process from chemical decoupling. Kinetic equilibrium between DM and the primordial plasma is maintained by rapid momentum exchange through scattering. Chemical equilibrium, on the other hand, holds when reactions that change the number of DM particles are active (for WIMPs, these reactions are typically annihilation processes). 
Kinetic equilibrium lasts much longer than chemical equilibrium due to the much higher number density of relativistic plasma particles available for scattering compared to the non-relativistic WIMPs necessary for annihilation. At chemical decoupling the annihilation/creation rates fall well below the Hubble expansion rate, and the dark matter particle number freezes out to a constant value per comoving volume. Still the dark matter remains in thermal equilibrium with the relativistic plasma through frequent elastic scattering. At  kinetic decoupling, even such scattering processes become slower than the Hubble expansion rate, and the dark matter becomes free from the heat bath and begins to stream freely. 
The dark matter kinetic decoupling determines the low-mass cutoff scale for the size of the dark matter halos (protohalos), which is of importance for structure formation in the Universe.




In this paper, we pay particular attention to the role of DM-quark interactions in the process of kinetic decoupling. The role of these interactions has not been fully explored (to the extent of our knowledge, only Ref.~\cite{bri5} includes them in the numerical analysis), contrary to the numerous studies that include scattering of DM and leptons.  
In addition to the lack of due attention to DM-quark interactions, another strong motivation for looking into the quark interactions stems from the unprecedented wealth of data from the Large Hadron Collider (LHC) and direct dark matter search experiments, which provide us with a direct probe of the DM-quark interactions. These particle physics experiments would in principle put useful astrophysical bounds on the size of dark matter protohalos.

The main results presented in this paper are (i) a kinetic-decoupling Fokker-Planck equation that extends the expression in Bertschinger \cite{bert} to general WIMP models (it was first obtained in the PhD thesis work of J. Kasahara under the direction of one of the authors (PG) \cite{kasahara}) and (ii) a connection between the smallest mass of the dark matter protohalos, particle searches at hadron colliders, and dark matter direct detection experiments (we illustrate this connection in the context of effective DM-quark interaction operators and of the Minimal Supersymmetric Standard Model).

The layout of the paper is as follows. First, in Sec. \ref{form}, we  outline how we estimate the kinetic decoupling temperature; here we generalize the formalism presented in Ref. \cite{bert} and give a general expression for the momentum exchange momentum relaxation rate that can be applied to any non-relativistic WIMP. Then in Sec. \ref{mi} we apply this  method to the DM-quark effective operators and show that the current collider and dark matter experiments set an upper bound on the smallest allowed protohalo mass. Finally in Sec. \ref{md} we quantify the importance of the DM-quark scattering relative to the DM-lepton scattering in the MSSM.

\section{The dark matter kinetic decoupling and the smallest protohalos}
\label{form}


In this section we outline the formalism to estimate the kinetic decoupling temperature and the corresponding smallest dark matter protohalo mass. 

Let us first start with a heuristic order of magnitude argument for the kinetic decoupling process before jumping to the Boltzmann and Fokker-Planck equations. The momentum transfer per collision between the plasma (heat bath) at temperature $T$ and the heavy dark matter particles $\chi$ of mass $m_\chi \gg T$ is of order $T$, much smaller than the average momentum $p$ of the dark matter particles, which is of order $(m_\chi T)^{1/2}$ as follows from the fact that the average kinetic energy of the dark matter particles, $p^2/(2m_\chi)$, is of order $T$. Many collisions, $\sim m_{\chi}/T$, are required for the dark matter to transfer a large part of its momentum to the plasma or to acquire it from the plasma. The momentum relaxation rate $\gamma$ is thus $\gamma \sim (T/m_{\chi}) \, \Gamma_{el} $, where $\Gamma_{el}$ is the elastic collision rate. The dark matter is in thermal equilibrium with the plasma when the momentum relaxation rate is larger than the Hubble expansion rate. Thermal decoupling occurs when the relaxation rate becomes of the order of the Hubble expansion rate, and this defines the kinetic decoupling temperature.

We can estimate the relaxation rate and consequently the decoupling temperature more accurately through the Fokker-Planck equation without approximating the dark matter as a perfect fluid or fully collisionless gas. Ref. \cite{kasahara} has re-derived the Fokker-Planck equation first discussed in Ref. \cite{bert} without assuming a specific cross section for WIMP-lepton scattering and allowing for massive particles in the plasma. The latter generalization is important at temperatures of the order of the electron or muon or quark masses. The former generalization goes beyond the zero-momentum transfer approximation of the previous literature \cite{brifirst,bri5,hofm,bere,wimpy}
(Mandelstam variable $t=0$), and allows the study of general particle models. Indeed the only assumptions in Ref. \cite{kasahara} are that the dark matter particle is heavy ($m_\chi \gg T $ or any other mass/energy scale) and that to momentum transfer is small. Under these assumptions, the Boltzmann equation for the time-dependence of the dark matter particle occupation number $f_\chi({\bf p}_\chi)$ in a Friedmann-Robertson-Walker universe reduces to the Fokker-Planck equation 
\ba
\frac{\partial f_{\chi}}{\partial t} - H {\bf p}_{\chi} \cdot \frac{\partial f_\chi}{\partial {\bf p}_\chi} =
\gamma(T)  \,\, \frac{\partial }{\partial {\bf p}_{\chi}}
\cdot \left(
{\bf p}_{\chi} f_{\chi}(1\pm f_\chi)+m_{\chi}T\frac{\partial f_{\chi}}{\partial {\bf p}_{\chi}}
\right)
\label{FP}
\ea
with momentum relaxation rate
\ba
\label{ourgamma}
\gamma(T) &=& \sum_{i} \frac{g_i}{6 m_{\chi} T} \int^{\infty}_{0} \frac{d^3{\bf p}}{(2\pi)^3} \, f_i \, (1\pm f_i) \, \frac{p}{\sqrt{p^2+m_i^2}}  \int^{0}_{-4p^2} dt \, (-t) \, \frac{d\sigma_{\chi+i\to\chi+i}}{dt} .
\label{eq:gamma}
\ea
Here the sum extends over the species $i$ in the relativistic plasma, with mass $m_i$, occupation number $f_i({\bf p}_i)$, and $g_i$ statistical degrees of freedom (e.g.\ $g_{q+\overline{q}}=12$ for the number of spin, color, and particle-antiparticle states of each flavor of Standard Model (SM) quark and antiquark). The + sign in $1\pm f_\chi$ and $1\pm f_i$ corresponds to bosons (stimulated emission), the - sign to fermions (Pauli blocking). 
And $d\sigma_{\chi+i\to\chi+i}/dt$ is the differential scattering cross section
 for the elastic scattering of $\chi$ and $i$, written as a function of the Mandelstam variable $t$ and of the center-of-mass momentum $p$, which in the heavy $\chi$ mass limit ($m_\chi \gg p$) equals the incoming momentum of particle $i$ in the plasma rest frame. Also, in the same limit,
\begin{align}
 \frac{d\sigma_{\chi+i\to\chi+i}}{dt}  = \frac{1}{64 \pi m_\chi^2 p^2} \, \overline{\left\vert{\cal M}_{\chi+i\to\chi+i}\right\vert^2},
\end{align}
where ${\cal M}_{\chi+i\to\chi+i}$ is the invariant scattering amplitude and an overline indicates the usual sum over final polarizations and average over initial polarizations.

The Fokker-Planck equation (\ref{FP}) automatically conserves the number of $\chi$ particles per comoving volume, since the time derivative of $a^3 f_\chi({\bf p}_\chi)$ is a total divergence in momentum space. The relaxation rate $\gamma(T)$ is an average over the thermal distribution of the mean square momentum transfer  $\overline{q^2} = \overline{-t}$ times the collision rate. 

The expression for the momentum relaxation rate $\gamma(T)$ in Eq. \ref{ourgamma} was first presented in Ref. \cite{kasahara} and it reproduces the formula for $\gamma(T)$ in Ref. \cite{bert} which only considered the bino scattering off massless leptons with a simplified invariant amplitude (the formula in Ref. \cite{brifirst,bri5}, which uses the forward scattering cross section $\left.d\sigma/dt\right|_{t=0}$ in place of $ (4p^2)^{-2} \int_{-4p^2}^{0} dt (-t) d\sigma/dt$ in Eq.~(\ref{eq:gamma}), gives a $\gamma(T)$ value which is 20\% larger). The formula given in Eq. \ref{ourgamma} can be applied to a generic scattering amplitude and is thus of wide use. It has been implemented \cite{kasahara} in an extension of the DarkSUSY computer code for particle dark matter \cite{darksusy}.

Multiplying the Fokker-Planck equation by the $\chi$ kinetic energy ${\bf p}_\chi^2/(2m_\chi)$, and integrating in $d^3 {\bf p}_\chi$ neglecting the stimulated emission or Pauli blocking factors ($1\pm f_\chi \simeq 1$), leads to an equation for the $\chi$ kinetic temperature $T_\chi$, defined as 2/3 of the average $\chi$ kinetic energy,
\begin{align}
\frac{d T_\chi}{dt} + 2  H T_\chi = - 2 \, \gamma(T) \, \left( T_\chi - T \right) ,
\label{eq:temp}
\end{align}
where $T$ is the plasma temperature. 
Refs. \cite{brifirst,bert} present analytic solutions for the case $\gamma(T)$ proportional to a power of $T$. At temperatures greater than the kinetic decoupling temperature $T_{kd}$, the $\chi$ particles are coupled to the plasma and $T_\chi \simeq T \propto a^{-1}$. At temperatures smaller than $T_{kd}$, $T_\chi \propto T^2 \propto a^{-2}$, as appropriate for non-relativistic particles of momenta ${\bf p}_\chi \propto a^{-1}$ that expand freely decoupled from the rest of the universe. 
In general a numerical solution of Eq. (\ref{eq:temp}) is necessary. Here we content ourselves with estimating 
the kinetic decoupling temperature $T_{kd}$ as the solution of \cite{bert,kasahara}
\ba
\label{cond}
\frac{\gamma (T_{kd})}{2}=H(T_{kd}) .
\ea

The kinetic decoupling temperature is important to establish the mass of the smallest dark matter protohalos. As pointed out in Refs. \cite{hofm,matias,bert}, acoustic oscillations due to the coupling between the dark matter and the plasma damp the amplitude of fluctuations at scales smaller than the horizon size at decoupling. After decoupling, dark matter particles can stream freely without interacting with the plasma, and this process erases fluctuations up to the distance to which they can stream from the time of kinetic decoupling. The mass of the smallest protohalo $M_{\rm halo,min}$ is determined by the larger of the DM mass inside the horizon at kinetic decoupling $M_{kd}$ and the DM mass within the free streaming length $M_{fs}$,
\begin{align}
M_{\rm halo,min} = \max ( M_{kd}, M_{fs} ) .
\end{align}

The DM mass contained within the free streaming length $\lambda_{fs}$ is
\ba
M_{fs} \approx \frac{4\pi}{3}\left( \frac{\pi}{k_*} \right)^3 \rho_{m0}
\ea
where $k_*=2 \pi/\lambda_{fs}$ and $\rho_{m0}$ is the dark matter density at the present time.
The comoving free streaming scale 
\ba
\lambda_{fs}=a_0\int_{t_{kd}}^{t_{0}} dt (v/a),
\ea
where $v \propto a^{-1}$ after kinetic decoupling, grows logarithmically during the radiation era ($a \propto t^{1/2}$) and saturates during the matter domination era ($a \propto t^{2/3}$). 
We find
(see Appendix)
\ba
\lambda_{fs} &=& \frac{v_{kd}a_{kd}}{2 a_{eq}C} \, \left\{ \left[\ln\frac{C\tau}{2+C\tau}\right]_{\tau_{kd}}^{\tau_0} + K_{fs} \right\}, 
\label{eq:lambdafs}
\ea
where
\ba
\frac{a_{kd}}{a_{eq}} = \frac{T_{eq}}{T_{kd}} \left( \frac{h(T_{eq})}{h(T_{kd})} \right)^{1/3},
\qquad
C \tau = \sqrt{1+\frac{T_{eq}}{T}}-1,
\qquad
C&\equiv& 
\frac{a_{eq}}{a_0} \sqrt{\frac{\pi G \rho_{eq}}{3}},
\ea
$\rho_{eq}$ 
is the total energy density at the time of matter-radiation equality. Finally, $K_{fs}$ is a correction term that takes into account the change in the effective number of degrees of freedom between kinetic decoupling and the time of matter-radiation equality,
\ba
K_{fs} &=& \int_{T_*}^{T_{kd}}\left[ \frac{\sqrt{g(T_*)}}{h^{2/3}(T_*)} \frac{h^{2/3}(T)}{\sqrt{g(T)}}  \left( 1+ \frac{1}{3} \frac{d\ln h(T)}{d\ln T} \right) - 1 \right] \, \frac{dT}{T} ,
\label{eq:kfs}
\ea
where $g(T)$ and $h(T)$ are the energy and entropy degrees of freedom, respectively, 
and $T_*$ is a temperature much smaller than the electron-positron annihilation temperature and much larger than the temperature at equality, $T_{eq} \ll T_* \ll 0.1 m_{e}$ (we take $T_*=1$ keV for concreteness (see Appendix for the details)). 
For the dark matter velocity at decoupling $v_{kd}$, we use $v_{kd}=\sqrt{6 T_{kd}/5 m_{\chi}}$, with the coefficient $6/5$ obtained numerically \cite{bert}. 

Acoustic damping is characterized by the dark matter mass inside the horizon at decoupling 
\ba
M_{kd}&\approx &\frac{4 \pi}{3}\frac{\rho_{m}(T_{kd})}{H^3 (T_{kd})} \\
&=&\frac{4 \pi}{3} \rho_{m0}  \frac{h(T_{kd})}{h(T_0)} \left( \frac{T_{kd}}{T_0}\right)^3  \frac{1}{{H^3}(T_{kd})},
\ea
where $T_0$ is the present temperature, $\rho_{m0}$ is the present DM density, and the Hubble parameter $H(T_{kd})$ at decoupling can be obtained from the Friedmann equation with total energy density $\rho(T_{kd}) = (\pi^2/30) g(T_{kd}) T_{kd}^4$ at kinetic decoupling. 
\begin{figure}
\begin{center}    
\epsfxsize = 0.6\textwidth
\epsffile{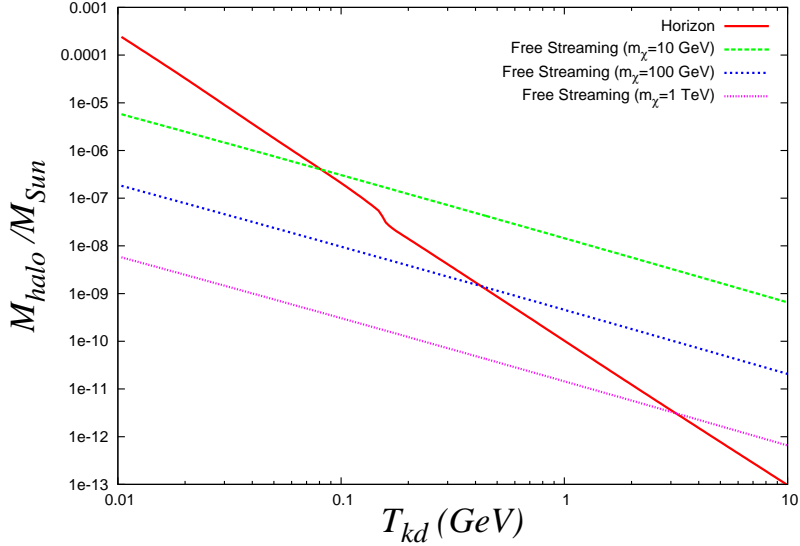}
\end{center}        
\caption{The smallest dark matter halo mass as a function the kinetic decoupling temperature $T_{kd}$. The smallest dark halo mass is the mass within the scale characterized by the larger between the acoustic damping length and the free streaming length: acoustic damping at relatively small $T_{kd}$, free-streaming at relatively large $T_{kd}$. The free streaming length depends on the dark matter particle mass $m_\chi$ through its velocity at kinetic decoupling, which scales as $\sqrt{T_{kd}/m_\chi}$. The feature around $T_{kd} \sim 150$ MeV is an imprint of the change in relativistic degrees of freedom during the QCD phase transition.
}
\label{hmass}
\end{figure}

For the effective energy and entropy degrees of freedom $g(T)$ and $h(T)$ we adopted the model of Ref. \cite{hp} (equation of state B) as implemented in DarkSUSY \cite{darksusy}. The factor $h(T_{kd})/h(T_{0})$, which was not present in Ref. \cite{bert}, takes into account the change of  comoving volume due to the entropy increase in the radiation. This ratio can be bigger than 10 for temperatures of the order of the QCD scale ($\gtrsim 200$ MeV in the model of our choice \cite{hp}). 

The acoustic damping and free streaming scales are plotted in Fig. \ref{hmass}. For a wide range of parameters, the acoustic damping scale is larger than the free streaming scale and thus determines the cutoff scale of the smallest halo size. The free streaming length becomes more important than the acoustic damping scale when $T_{kd} /  m_{\chi}$ becomes large as seen in the figure. This behavior is expected because the acoustic damping length scales as $\tau_{kd} \sim 1/T_{kd}$, while the free streaming length roughly scales as $\sqrt{T_{kd}/m_{\chi}}\tau_{kd} $. Notice that the scale of the smallest dark protohalo decreases with increasing kinetic decoupling temperatures.


\section{Quark interactions in the effective field theory}
\label{mi}
It is illustrative to consider the effective field theory approach to see the essential features of the DM-quark interactions and their relations to the kinetic decoupling temperature. A great feature of studying such effective DM-quark operators is that one can study the range of the kinetic decoupling temperature allowed by the recent data from the LHC and dark matter direct detection experiments that directly probe DM-quark interactions. 

The effective field theory approach has been studied extensively and we refer the readers to the existing literature for a complete survey of the effective operators \cite{russ,good10b,good10c,cheu,kilic,foxlhc,foxtev,ibe,belt,tim11,liam} \footnote{We do not consider the UV completion of the effective operators of our interest. Strictly speaking the effective theory breaks down if the 4-momentum transfer is comparable to or larger than the mass of a particle mediating the interaction (the typical momentum transfer is of order $T_{kd}$ for  DM kinetic decoupling). See for instance Refs. \cite{foxtev,russ} for further discussions on the validity of the effective operator approach in the case of light mediators.}. Because our purpose in this paper is to study the potential significance of the DM-quark scattering in the dark matter kinetic decoupling processes, which can be constrained from the current LHC and dark matter experiment data, we simply assume the dark matter is a Majorana fermion and a SM singlet, and consider the following scalar and axial-vector effective DM-quark point-interaction operators relevant for the direct dark matter search experiments,
\ba
{\cal O}_S&=&\sum _q \frac{m_q}{\Lambda^3} \bar{\chi} \chi \bar{q} q , \\
{\cal O}_A&=&\sum_q
\frac{1}{\Lambda^2} (\bar{\chi} \gamma^{\mu} \gamma^5 \chi) (\bar{q} \gamma_{\mu} \gamma^5 q) .
\ea 
These operators lead respectively to spin-independent and spin-dependent interaction whose interaction strength is set by the effective cutoff scale $\Lambda$. The other DM-quark interaction operators besides the scalar and axial-vector operators vanish in the non-relativistic limit for Majorana fermion dark matter.
 
We make the simplifying assumptions that the DM  couples universally to the Standard Model (SM) up and down type quarks via $\cal{O}_A$, and that it couples in $\cal{O}_S$ through the quark mass suppression factor $m_q$ implied by chirality breaking. 

The spin-independent cross section per proton reads 
\ba
\sigma_{SI}
= \frac{4 \mu_p^2 m_{p}^2 f_{m,\rm eff}^2}{\pi \Lambda^6}
\ea
where $\mu_p=m_pm_\chi/(m_p+m_\chi)$ is the proton-DM reduced mass and $f_{m,\rm eff}$ is the fraction of the proton mass $m_p$ coupled to the scalar operator $\cal{O}_S$, which is equal to the mass fraction carried by quarks plus 6/27 of the mass fraction carried by gluons (for the default values of the nucleon parameters in DarkSUSY, which we use, $f_{m, \rm eff}=0.375$).
The spin-dependent cross section per nucleon is
\ba
\sigma_{SD} = \frac{16\mu_p^2}{\pi \Lambda^4} \left(\sum_q \Delta q \right)^2 J(J+1) ,
\ea
where we use the value $\sum_q \Delta_q = 0.32$ for the spin fraction carried by quarks in the nucleon \cite{keith3,compass} and $J$ is the nuclear spin (1/2 for a free nucleon). 
 For the dark matter direct search experiment constraints, we use the data from SIMPLE \cite{simple} for the spin-dependent cross section and from XENON100 \cite{xenon100} for the spin-independent interaction, which are currently the most sensitive direct detection experiments at the high end of the DM mass. 

\begin{figure}[ht]
\centering
\subfigure[~~Scalar effective operator]{
\includegraphics[scale=0.675]{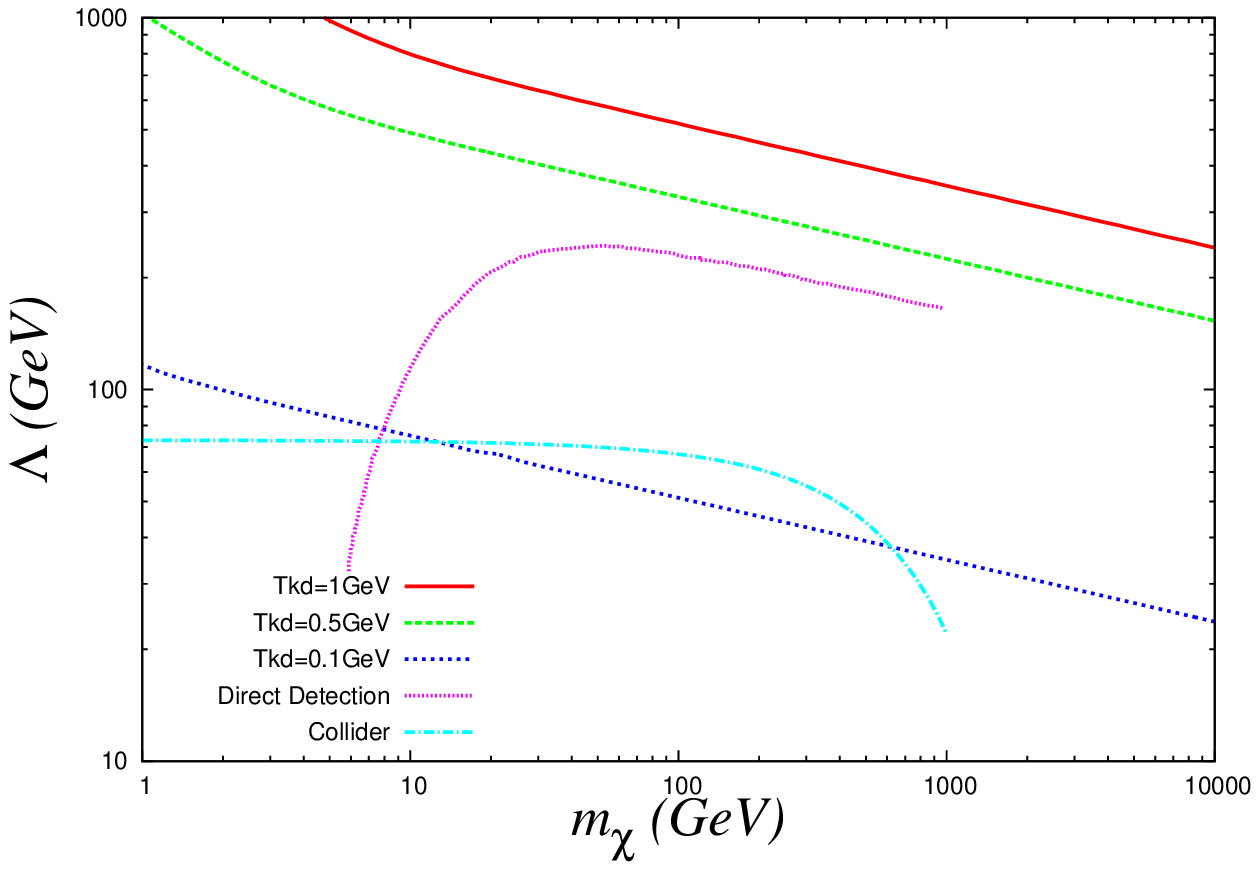}
}
\subfigure[~~Axial-vector effective operator]{
\includegraphics[scale=0.675]{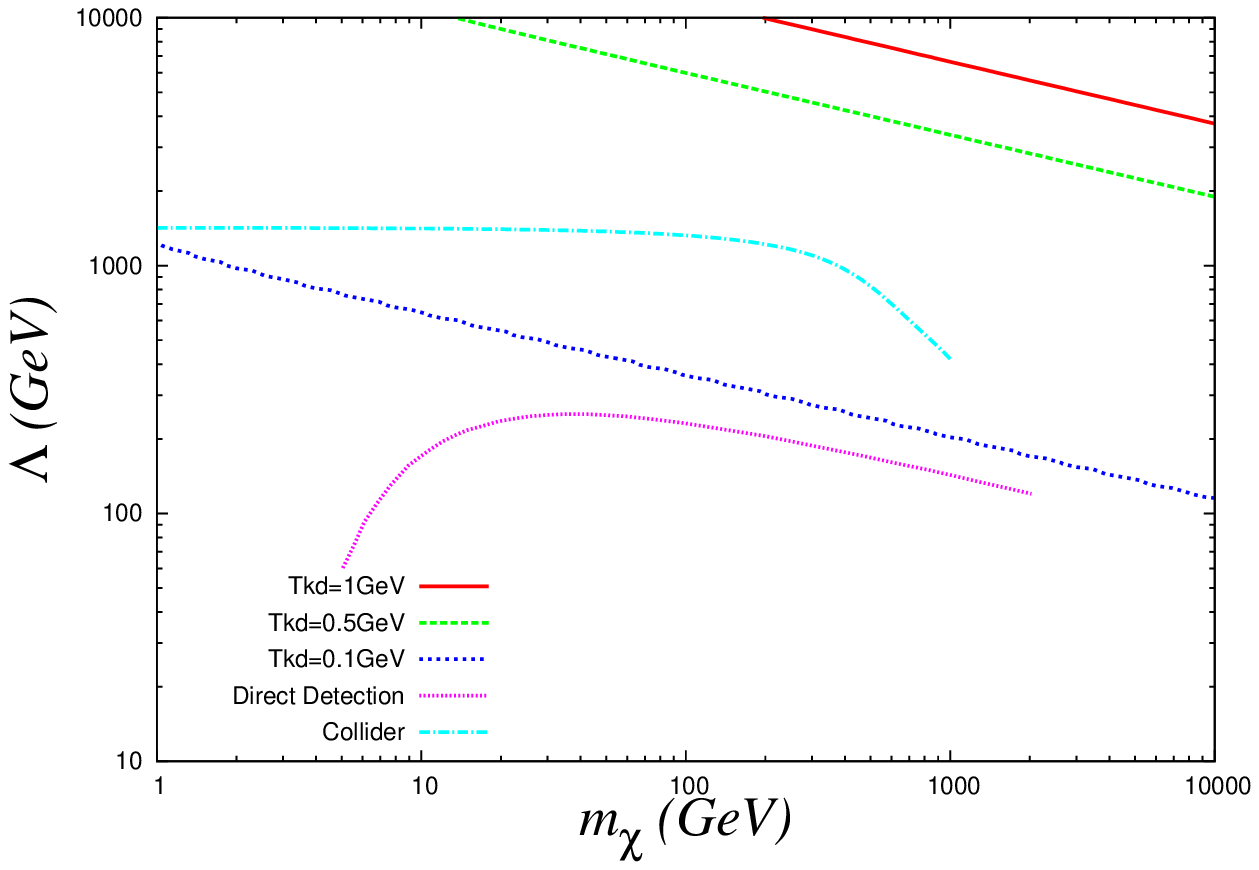}
}
\label{fig:directlhc}
\caption[Optional caption for list of figures]{
Bounds on the effective interaction scale $\Lambda$ as a function of the dark matter mass $m_\chi$ for the scalar (left subfigure) and axial-vector (right subfigure) point-interaction effective operator. Direct detection and collider experiments exclude the regions below their respective lines (densely-dotted and dashed-dotted). These bounds become weaker at small and large $m_\chi$ respectively, due to a minimum detectable energy in direct detection experiments and a maximum beam energy in collider experiments. The large $m_\chi$ decrease of the direct detection bounds is due to the $m_\chi^{-1}$ scaling of the dark matter flux onto the detector. We see that the combination of direct detection and collider bounds forces the kinetic decoupling temperature to be larger than $\sim 100$ MeV, a regime in which DM-quark scattering can be important.

}
\end{figure}

A complementary set of constraints comes from collider experiments. 
Collider constraints do not suffer from the astrophysical uncertainties, such as the local galactic dark matter density and velocity distribution, that afflict direct search experiments. 
The recent LHC data release from the CMS Collaboration \cite{cms47} presented 1142 observed events in a mono jet analysis with leading jet transverse momentum $p_T>110$ GeV, pseudorapidity $|\eta|<$2.4 and missing transverse energy $\not\!\!E_t>350$ GeV, to be compared with a Standard Model prediction of $1224 \pm 101$ for the data sample of 4.7/fb total integrated luminosity at a center-of-mass energy of 7 TeV. We set a 2$\sigma$ collider lower bound on the effective coupling scale $\Lambda$. 
We implement the effective operators by treating the scalar and axial-vector operators separately (one at a time) in Madgraph/Madevent.  We use Pythia for the hadronization and the initial/final state radiation, treating the jets with the pycell subroutine. 

The lower bounds on $\Lambda$ from the direct dark matter search and collider experiments are shown in Fig. 2 as a  function of the dark matter mass $m_\chi$, along with contours of the kinetic decoupling temperature $T_{kd}$. The LHC sensitivity decreases toward  higher dark matter masses due to kinematic reasons. The collider bounds show the limits for the mono-jet events without additional nearby jets, in accord with our simplifying assumption of perfect efficiency. 
The direct detection experiments on the other hand suffer from the threshold to detect the DM recoil, and consequently, for the scalar operator, the collider constraints become stronger for light DM ($m_\chi \lesssim $ 10 GeV). 
The spin-dependent DM-nucleus scattering is not enhanced by the square of the atomic mass of the target nucleus, in contrast to the spin-independent interaction rate, and hence direct search experiments lead to bounds on the scale of the axial-vector operator that are weaker than the current LHC constraints .

The combination of direct detection and collider bounds in Fig. 2 forces the kinetic decoupling temperature to be larger than $\sim100$ MeV, a regime in which DM-quark scattering becomes important.
 The power-law behavior of the kinetic decoupling temperature contours in the figure emerges from the relation $H(T_{kd})=\gamma(T_{kd})/2$ and power-law dependence of the low-momentum transfer relaxation rate $\gamma(T)$ on $\Lambda$, $T$ and $m_{\chi}$.


Our effective-operator analysis provides us with upper bounds on the smallest dark protohalo mass directly from the current LHC and direct dark matter search limits, without having to scan the parameters space of a specific particle model. For instance, 
we find that for $m_{\chi}=300$ GeV the smallest allowed kinetic decoupling temperature is 350 MeV for the scalar operator and 150 MeV for the axial-vector operator, corresponding to upper limits on the smallest protohalo mass of $3 \times 10^{-9} M_{\Sun}$ and
 $5 \times 10^{-8} M_{\Sun}$ respectively. Note that here the protohalo cutoff mass scales as $M_{\rm halo,min}\propto (T_{kd}\sqrt{g_{eff}(T_{kd})})^{-3} h_{eff}(T_{kd})$. Independently of the DM mass and spin-dependent or -independent interaction, there is an absolute lower bounds on the kinetic decoupling temperature of $\sim 100$ MeV and a corresponding absolute upper bound of $\sim 10^{-6} M_{\Sun}$ (of order of the Earth's mass) on the mass of the smallest dark protohalos.

In the model-independent analysis of this section, each quark was decoupled at its mass scale but the QCD phase transition was not taken account of which is currently heavily model dependent. Indeed our simplified analysis adding quarks to the plasma as a free gas is not really accurate during the QCD phase transition, when hadrons are also present, whose treatment would require more input from lattice simulations. Nonetheless, our findings show that the current LHC and direct dark matter search data impose lower bounds on \tkd that are well in the QCD phase transition regime, supporting our statement that DM-quark interactions, hitherto neglected in studies of kinetic decoupling, are important and must be included. The inclusion of DM-quark interactions will become even more important as forthcoming LHC/dark matter experiments probe DM-quark interactions further, potentially pushing the lower bound on \tkd up and above the QCD transition regime. 




\section{Quark interactions in a concrete model}
\label{md}
An advantage of specifying a concrete model rather than effective DM-quark operators is that in a concrete model all the DM interaction terms are present, including the interference terms, which are cumbersome to specify in a model independent approach treating the different effective operators separately. In this section, we use a specific model to examine the relative significance of DM-quark interactions in comparison to DM interactions with other particles.

As a concrete model, we choose the Minimal Supersymmetric Standard Model (MSSM), in which at the varying of the model parameters the values of \tkd are known to span a wide range, from tens of MeV to more than a GeV \cite{wider,chen,wimpy,hofm,kam2,brifirst,bert,bri5,torsten3,stefano,bi,matias,boe,bere}. Because of this, the MSSM provides a good theoretical proving ground to check the relative importance of the DM-quark interactions compared with the DM scattering off leptons that has been discussed extensively in the previous literature. 

The elastic scattering of neutralinos off the fermions in the MSSM occurs through the exchange of gauge bosons, Higgs bosons, and sfermions.
We extended the modified DarkSUSY first developed in Ref. \cite{kasahara} so as to include all possible dark matter interactions and interference terms in the MSSM for dark matter scattering off fermions in the primordial plasma.
For the effective number of degrees of freedom, we adopt the model of Ref. \cite{hp} with QCD phase transition temperature of 154 MeV (equation of state B \cite{hp}, the default model in DarkSUSY \cite{darksusy}). For the scattering of DM off quarks, the Fokker-Planck equation is numerically fully solved down to 154 MeV including all the relevant interactions and we simply turn off the DM-quark scatterings below 154 MeV. Even though the temperature at which the asymptotically free quark description becomes valid is expected to be higher than the QCD phase transition temperature, this simplification suffices for our purpose of showing the potential significance of DM-quark scattering in estimating the dark matter protohalo mass. Choosing a higher temperature for the threshold of DM-quark scattering to ensure the validity of the free quark description does not change the conclusions of this section (adding DM-quark scattering besides DM-lepton scattering can increase the protohalo mass estimation by a factor of 2 or more).


We scanned the MSSM-7 parameter space, characterized \cite{bp2} by seven parameters specified at the electroweak scale: 2 trilinear A-terms, a soft sfermion mass parameter, a gaugino mass parameter, and three Higgs-sector parameters. The trilinear A-terms and the soft sfermion mass matrices were assumed to be diagonal to avoid flavor-changing-neutral-current issues and they were parameterized as ${\bf A}_U=diag(0,0,A_t)$, ${\bf A}_D=diag(0,0,A_b)$. All the soft sfermion mass matrices are $m_0$ times the identity matrix. The gaugino mass GUT relation was assumed and the SU(2) gaugino mass $M_2$ was chosen as the free parameter. The Higgs sector was parameterized by the CP-odd Higgs boson mass $M_A$, the Higgsino mass parameter $\mu$ and the ratio $\tan \beta$ of Higgs vacuum expectation values. These seven parameters were randomly scanned over the ranges $10 \leq \tan \beta \leq 50$, $0 \leq |A_{t,b}|  \leq 10$ TeV and $[10 GeV, 10 TeV]$ for the remaining mass parameters $m_0$, $M_2$, $M_A$, and $\mu$. We applied the phenomenological bounds available in the DarkSUSY subroutines \cite{darksusy} for the theoretical consistency of the model and the experimental constraints. We excluded parameter sets that have charged or colored vacua, that have potentials unbounded from below, that do not have the neutralino as the lightest supersymmetric particle, and that violate experimental bounds on supersymmetric masses or rare processes such as $b\rightarrow s \gamma$. 

Because our purpose here is to show quantitatively the significance of the DM-quark scattering
we do not require that the neutralino thermal relic abundance should match the observed cold dark matter value (actually, through non-thermal production, entropy production, or non-standard expansion history, the neutralino thermal relic abundance can be smaller or larger than the cosmological value $\Omega_{DM}\approx 0.1$, even if the neutralinos comprise all of the cold dark matter \cite{paologelmini}). 



\begin{figure}[htb!]
\begin{center}    
\epsfxsize = 0.6\textwidth
\epsffile{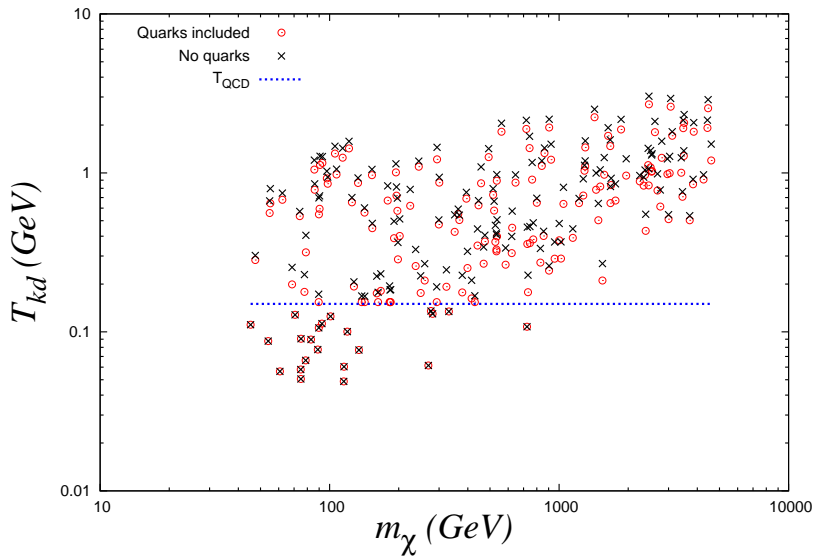}
\end{center}        
\caption{The neutralino kinetic decoupling temperature as a function of the neutralino mass for a random scan in the MSSM-7 parameter space. The inclusion of DM-quark scattering  can change the kinetic decoupling temperature by 30\% and hence the dark matter protohalo mass by more than a factor of 2.
}
\label{mssm7}
\end{figure}

In Fig.~\ref{mssm7} we compare the neutralino kinetic decoupling temperatures with and without including DM-quark interactions for a random scan in the MSSM-7 parameter space. The open circles represent the \tkd values computed including the DM-quark interactions, and the crosses are the \tkd values computed without including the DM-quark interactions. 
We first observe that kinetic decoupling can occur above the QCD phase transition scale, as first pointed out in Ref. \cite{kam2}, which performed an analogous numerical scan without including the DM-quark interactions.  
For $T_{kd} \lesssim T_{QCD}$, the quarks are bounded inside the hadrons and the pions were much less abundant than the light leptons in the thermal bath. We numerically checked that the inclusion of pion scattering affects the kinetic decoupling temperature by less than one per cent in our scenarios \footnote{The leading order DM-pion coupling term $c \pi^2 \bar{\chi} \chi$ can be estimated with a coupling constant \cite{kami1}
\ba
c= \frac{m_{\pi}^2}{2(m_u+m_d)}(f_u+f_d),
\ea
where $f_q$, with $q=$ up or down quark in this case, is the coefficient of the effective scalar current operator $f_q \bar{q}q \bar{\chi}\chi$ in the MSSM \cite{mihoko}. This form of $c \pi^2 \bar{\chi} \chi$ was derived using the soft-pion technique in the framework of chiral perturbation theory, which is valid for an energy scale $\lesssim 4 \pi f_{\pi}$ ($f_{\pi}$ is the pion decay constant), and hence is applicable to the soft-pion case.}.
Fig. \ref{mssm7} shows that \tkd can be significantly influenced by  DM-quark interactions for \tkd above the QCD scale. In our MSSM-7 parameter scan we found that the typical ratio of \tkd with and without DM-quark interactions ranged between 1 and $\sim1.3$.
When the smallest protohalo mass $M_{\rm halo,min}$ is set by the acoustic damping scale, it scales with \tkd as $M_{\rm halo,min}\sim T_{kd}^{-3}$. Hence the inclusion of DM-quark scattering can lead to a factor 2 or more increase in the smallest protohalo mass for a wide range of MSSM parameter values. When the smallest protohalo mass $M_{\rm halo,min}$ is set by the free streaming length, it roughly scales as $M_{\rm halo,min}\sim T_{kd}^{-3/2}$, and it less affected by the inclusion of DM-quark interactions.
  
This conclusion on the importance of including DM-quark interactions would not change even if we had switched off DM-quark scattering up to temperatures four or five times higher than the QCD phase transition temperature, as to guarantee the presence of free quarks, because our MSSM-7 parameter scan already covers such regime of \tkd well above the QCD phase transition temperature \footnote{That such high \tkd are common can be seen for instance in the MSSM scans in Ref. \cite{bri5}, which used a different QCD phase transition model and a different method of estimating the kinetic decoupling temperature.}.


\section{Discussion/Conclusion}
Studies of particle dark matter can provide us with a unique link between particle physics and astrophysics. To illustrate one such connection, we examined the importance of DM-quark interactions in the kinetic decoupling of particle dark matter and the consequent mass of the smallest dark protohalos. We focus on a model independent analysis through effective DM-quark interaction operators and a model specific analysis within a seven-parameter MSSM. 

In the effective operator approach, in which we assumed DM-lepton interactions are negligible, we found that current direct dark matter search and collider constraints force the DM kinetic decoupling temperature $T_{kd}$ to exceed $\sim100$ MeV, whether the interaction is spin-dependent or spin-independent. This sets an upper limit of $\sim10^{-6} M_{\Sun}$ (of order of the Earth's mass) to the mass of the smallest dark protohalo. 

In the MSSM-7 study, we found that inclusion of neutralino-quark interactions can increase the smallest dark protohalo mass by more than a factor of 2 whenever the neutralino decoupling temperature exceeds the QCD transition temperature and acoustic damping dominates over free streaming.

The constraints on the kinetic decoupling temperature from the current LHC and direct dark matter search experiments turned out to be in the regime of the QCD phase transition. The forthcoming LHC and direct dark matter search data will most likely push this \tkd bound up, possibly beyond the quark-hadron transition, into a regime in which our approximation of a gas of free quarks is definitely applicable.

If DM-quark interactions are discovered at the LHC or in direct detection experiments, measurements of their strength can provide us with a lower bound of the mass of the smallest dark matter protohalos, once proper account is taken of  possible additional interactions with leptons.

The survival of the smallest dark matter protohalos to the present time, and their observability, depend strongly on complicated tidal forces and other astrophysics such as stellar interactions, and have been the subject of vigorous debate. More input from computer simulations of structure formation in the universe would help clarify these issues, despite the vast dynamic range of masses and non-linear effects involved in simulating the relevant processes \cite{die,kra,gne,sava,bere5,gre,ost}. 

Although we discussed only the hadron collider (LHC) data that constrain the DM-quark interaction operators, an analogous exercise can be performed for the DM-lepton interactions considering a lepton collider \cite{foxlep,mam} which is crucial for a concrete model such as the MSSM where DM-lepton interactions cannot be neglected for the estimation of the kinetic decoupling. Inclusion of non-collider experiments and indirect search experiments also deserve further studies. For instance the constraints from the cosmic antiproton flux can put relatively tight constraints on the axial-vector operator \cite{adri,che}, and more systematic studies of the dark matter kinetic decoupling including additional experiments and additional effective operators beyond those considered here are worthy of future work. 

Our results on the impact of collider and dark matter search experiments onto the formation of structure in the early universe calls for further exploration of the creation and evolution of dark matter protohalos, in view of their potential role in probing the nature of the dark matter.


\section*{Acknowledgments}
For the numerical estimation of kinetic decoupling temperature, we have modified the DarkSUSY where the momentum relaxation rate and the full complement of supersymmetric scattering cross sections were first implemented by Junya Kasahara for his PhD thesis project \cite{kasahara}.
This work was supported in part by a Grant-in-Aid for the Global COE program and for Scientific Research from the MEXT of Japan (JH and KK), and by NSF award PHY-1068111 (P.G.). We thank the Kavli Institute for Theoretical Physics China where part of this work was conducted. P.G. also thanks the Korean Institute for Advanced Studies, Nagoya University and Seoul National University for support and hospitality during the completion of this work.
\appendix

\section{Derivation of the free streaming length formula}

Here we sketch the derivation of Eq.~(\protect\ref{eq:lambdafs}) for the free streaming length, which includes the change in the effective degrees of freedom before matter-radiation equality. Since the expressions for the energy and entropy degrees of freedom $g(T)$ and $h(T)$ usually do not extend down to temperatures $\sim T_{eq}$, we divide the integration in
\begin{align}
\lambda_{fs} = a_0 \int_{t_{kd}}^{t_*} dt \, \frac{v}{a} 
\end{align}
into two parts, which we join at a temperature $T_*$ defined so that $aT={\rm const}$ for $T<T_*$. The temperature $T_* $ is thus much smaller than the electron-positron annihilation temperature but much larger than the temperature at equality.

For $T<T_*$, we use an FRW model with matter and radiation, 
\begin{align}
H^2(a)=\frac{H_{eq}^2}{2} \left[ \left(\frac{a}{a_{eq}}\right)^{-3} +  \left(\frac{a}{a_{eq}}\right)^{-4} \right],
\end{align}
where $H_{eq} = \sqrt{8\pi G \rho_{eq}/3}$ is the Hubble parameter at the matter-radiation equality ($a=a_{eq}$), with $\rho_{eq}$ equal to the total density at that time (the contribution to $\lambda_{fs}$ from the low-redshift cosmological constant term is negligible). We find
\begin{align}
a_0 \int_{t_*}^{t_0} dt \, \frac{v}{a} & = a_0 v_{kd} a_{kd} \int_{t_*}^{t_0} \frac{dt}{a^2} \\
& =  a_0 v_{kd} a_{kd} \int_{a_*}^{a_0} \frac{da}{a^3 H(a)} \\
& = \sqrt{2}  \frac{a_0 v_{kd} a_{kd}}{a_{eq}^2 H_{eq}} \left[ 
\ln \frac{\sqrt{1+\alpha}-1}{\sqrt{1+\alpha}+1} \right]_{\alpha=a_*/a_{eq}}^{\alpha=a_0/a_{eq}} \\
& = \frac{v_{kd} a_{kd}}{2 a_{eq} C} \left[ \ln\frac{C\tau}{2+C\tau} \right]_{\tau_*}^{\tau_0},
\label{eq:lambdafs1}
\end{align}
where, on using $aT={\rm const}$ for $a_*<a<a_0$,
\begin{align}
C\tau &= \sqrt{1+\frac{T_{eq}}{T}} - 1, 
\end{align}
with
\ba
C\equiv 
\frac{a_{eq} H_{eq}}{a_0\sqrt{8}} =
\frac{a_{eq}}{a_0} \sqrt{\frac{\pi G \rho_{eq}}{3}}.
\ea

For $T>T_*$, we include the full temperature dependence of the degrees of freedom. Using conservation of total entropy $a^3 T^3 h(T) = {\rm const}$, from which
\begin{align}
\frac{da}{a} = - \left( 1 + \frac{1}{3} \frac{d\ln h(T)}{d \ln T} \right) \, \frac{dT}{T} ,
\end{align}
we find
\begin{align}
a_0 \int_{t_{kd}}^{t_*} dt \, \frac{v}{a} 
& =  a_0 v_{kd} a_{kd} \int_{a_{kd}}^{a_*} \frac{da}{a^3 H(a)} \\
& =  \frac{a_0 v_{kd} a_{kd}}{a_*^2 H(a_*)} \int_{a_{kd}}^{a_*} \frac{a_*^2 H(a_*)}{a^2 H(a)} \frac{da}{a} \\
& = \frac{v_{kd} a_{kd}}{2 a_{eq} C} \frac{1}{\sqrt{1+\frac{T_{eq}}{T_*}}} \int_{T_*}^{T_{kd}}  \frac{\sqrt{g(T_*)}}{h^{2/3}(T_*)} \frac{h^{2/3}(T)}{\sqrt{g(T)}}   \left( 1 + \frac{1}{3} \frac{d\ln h(T)}{d \ln T} \right) \, \frac{dT}{T}.
\label{eq:lambdafs2}
\end{align}
In the last step, we wrote $H(T_*)$ in terms of $H_{eq}$, and then $C$, using the matter and radiation model at $T<T_*$ where $aT={\rm const}$,
\begin{align}
a_*^2 H(T_*) = a_{eq}^2 \frac{H_{eq}}{\sqrt{2}} \sqrt{1+\frac{a_*}{a_{eq}}} .
\end{align}

After adding Eqs.~(\ref{eq:lambdafs1}) and (\ref{eq:lambdafs2}), we formally extend the integration in Eq.~(\ref{eq:lambdafs1}) from $\tau_{*}$ to $\tau_{kd}$ using the definition
\begin{align}
C\tau = \sqrt{1+\frac{T_{eq}}{T}} -1
\label{eq:formaltau}
\end{align}
and the mathematical identity
\begin{align}
 \left[ \ln\frac{C\tau}{2+C\tau} \right]_{\tau_{kd}}^{\tau_*} = \int_{T_*}^{T_{kd}} \frac{1}{\sqrt{1+\frac{T_{eq}}{T}}} \frac{dT}{T} .
\end{align}
Notice that  since $aT\ne {\rm const}$ for $T>T_*$, the variable $\tau$ in Eq.~(\ref{eq:formaltau}) is not the conformal time at plasma temperatures $T>T_*$, as instead it is at $T<T_*$, but is only a convenient mathematical variable. This gives
\begin{align}
\lambda_{fs} =  \frac{v_{kd} a_{kd}}{2 a_{eq} C} \left\{ \left[ \ln\frac{C\tau}{2+C\tau} \right]_{\tau_{kd}}^{\tau_0} + K_{fs} \right\},
\end{align}
with
\begin{align}
K_{fs} = \frac{1}{\sqrt{1+\frac{T_{eq}}{T_*}}} \int_{T_*}^{T_{kd}}  \frac{\sqrt{g(T_*)}}{h^{2/3}(T_*)} \frac{h^{2/3}(T)}{\sqrt{g(T)}}   \left( 1 + \frac{1}{3} \frac{d\ln h(T)}{d \ln T} \right) \, \frac{dT}{T} - \int_{T_*}^{T_{kd}} \frac{1}{\sqrt{1+\frac{T_{eq}}{T}}} \frac{dT}{T} .
\end{align}
The approximation $T_* \gg T_{eq}$ then gives the expression of $K_{fs}$ in Eq.~(\ref{eq:kfs}).

\end{document}